\documentclass[12pt]{article}
\usepackage[margin=1in]{geometry}
\usepackage{pstricks,pst-node,pst-text}
\usepackage{amsthm}
\usepackage{amssymb}
\usepackage{amsmath}
\theoremstyle{remark}
\DeclareMathOperator{\Tr}{Tr}
\DeclareMathOperator{\Hessian}{Hess}
\usepackage{amsfonts}
\usepackage{enumerate}
\usepackage{verbatim}
\usepackage[english]{babel}
\usepackage{graphicx}
\usepackage[cp1250]{inputenc}
\usepackage{natbib}
\usepackage{hyperref}
\usepackage{booktabs}
\usepackage{authblk}
\definecolor{alizarin}{rgb}{0.82, 0.1, 0.26}

\numberwithin{equation}{section}

\newtheorem{rem}{Remark}[section]

\newcommand{\beqo}{\begin{eqnarray*}}
\newcommand{\eeqo}{\end{eqnarray*}\noindent}
\newcommand{\beq}{\begin{eqnarray}}
\newcommand{\eeq}{\end{eqnarray}\noindent}

\def\be{{\mathbb{E}}}

\def\bq{{\mathbb{Q}}}

\begin{document}

\title{\LARGE \textbf{Pricing equity-linked life insurance contracts with multiple risk factors by neural networks}}
\date{Version: \today }

\author[1]{\large Karim Barigou\thanks{Karim.Barigou@univ-lyon1.fr (Corresponding author)}}
\author[2] {\large {\L}ukasz Delong\thanks{lukasz.delong@sgh.waw.pl}}
\affil[1]{\footnotesize ISFA, Univ Lyon 1, UCBL, LSAF EA2429, F-69007, Lyon, France}
\affil[2]{\footnotesize SGH Warsaw School of Economics, Collegium of Economic Analysis, Institute of Econometrics, Niepodleg{\l}o\'{s}ci 162, Warsaw 02-554, Poland}

\maketitle

\begin{abstract}

This paper considers the pricing of equity-linked life insurance contracts with death and survival benefits in a general model with multiple stochastic risk factors: interest rate, equity, volatility, unsystematic and systematic mortality. We price the equity-linked contracts by assuming that the insurer hedges the risks to reduce the local variance of the net asset value process and requires a compensation for the non-hedgeable part of the liability in the form of an instantaneous standard deviation risk margin. The price can then be expressed as the solution of a system of non-linear partial differential equations. We reformulate the problem as a backward stochastic differential equation with jumps and solve it numerically by the use of efficient neural networks. Sensitivity analysis is performed with respect to initial parameters and an analysis of the accuracy of the approximation of the true price with our neural networks is provided.
	
	
\end{abstract}

\newpage

\section{Introduction}

An equity-linked life insurance contract is a contract between a policyholder and an insurance company which presents insurance and investment features (\cite{palmer2006equity}, \cite{bauer2008universal}). Typically, the
policyholder pays either a single premium or a stream of periodic premiums
during an accumulation phase. In return, the insurer guarantees a stream of
periodic payments starting either immediately or at a future date. For
investors, equity-linked life insurance contracts allow to participate in the equity market and offer higher
returns than fixed annuity contracts. Moreover, they offer
protection against downside shocks in the financial market by the presence of diverse
guarantees. In the context of increasing life expectancies and the
population facing the need for sustainable income, the demand for guaranteed
income after retirement offered by equity-linked contracts is important (\cite{haefeli2013variable}).

Pricing of equity-linked life insurance contracts is a classical theme in the actuarial literature. Typically, this involves the choice of a financial and actuarial model to capture the financial and insurance risks. Based on no-arbitrage arguments and using the standard assumption that the financial and insurance risks are independent, these two components are merged together by a product of risk-neutral expectations (see \cite{fung2014systematic},
\cite{da2015valuing}, \cite{ignatieva2016pricing} among others). However, this approach does not address the fundamental issue of incompleteness arising from the insurance and the financial market. Since the risks cannot be perfectly hedged, the valuation is non-linear and we should proceed in two-steps by disentangling the hedgeable and non-hedgeable part of the liability (\cite{pelsser2014time}, \cite{dhaene2017fair} and \cite{barigou2019fair}).

We follow the pricing approach from \cite{delongfvtheory} where the authors derived a system of partial differential equations (PDEs) for the fair valuation of a stream of payments contingent on financial and insurance risk factors. In \cite{delongfvtheory}, three risk factors were considered and numerical methods for solving the PDEs were not discussed. In practice, we would like to apply a valuation operator which takes into account multiple correlated risk factors. In this paper we deal with six stochastic financial and insurance risk factors: interest rate (two factors), equity, volatility, unsystematic and systematic mortality. The interest rate is modelled with two-factor Hull-White model, the equity is modelled with Heston model where the volatility is modelled with Cox-Ingersoll-Ross model, the lifetimes of policyholders are modelled, conditionally on the force of mortality, with independent exponential distributions and the force of mortality is modelled with Feller process. We assume that equity-linked contracts include two types of minimum guarantees: death benefits in case of the deaths of the policyholder and survival benefits in case the policyholder is still alive at the contract termination. In such a framework, we can derive a system of PDEs for the fair valuation of a portfolio of equity-linked contracts. Solving such a system of PDEs seems to be challenging due to a large number of state variables and a large number of PDEs which reflects the number of policies in the insurance portfolio. It is known that finite difference methods for PDEs in high dimensions are not feasible due to the curse of dimensionality. In our case, we face an additional difficulty if we valuate a large portfolio of equity-linked contracts, since we have to solve recursively a large number of PDEs, where the solution of one equation (for the number of policyholders equal to $n$) is used as the input to the next equation (for the number of policyholders equal to $n+1$).

In recent years, some papers have considered the pricing of equity-linked contracts by neural networks. \cite{doyle2019using} explored the use of neural networks to price and hedge equity-linked contracts. They showed that neural networks offer an important computational gain compared to crude Monte-Carle methods, however they did not consider stochastic mortality neither stochastic volatility in their framework. In \cite{hejazi2016neural}, the authors proposed a neural network approach to price equity-linked contracts for large insurance portfolios. Their framework prices some representative contracts via Monte-Carlo and uses a spatial interpolation method to price the whole portfolio.

Another stream of literature investigated the use of neural networks to accelerate PDE solvers. \cite{han2018solving}, \cite{weinan2017deep}, \cite{beck2019machine} take advantage of reinforcement learning to speed up solving high-dimensional partial differential equations by using a so-called \textit{Deep BSDE} method. Moreover, \cite{chan2018machine} proposed an improved version of the Deep BSDE algorithm using different network architectures and parameterizations.

We see three contributions of this paper. First, we consider the pricing of equity-linked contracts in a general incomplete financial-actuarial market with stochastic interest rate, equity, volatility and mortality. In the spirit of \cite{delongfvtheory}, we derive a system of multi-dimensional non-linear PDEs and the optimal hedging strategy for our fair valuation problem – this is our analytical contribution. Second, we present a numerical contribution: using the representation of the PDEs as a BSDE with jumps, we show how to solve numerically the pricing problem by neural networks. In particular, we show how the jump component in the BSDE, which reflects the number of PDEs in the system of equations and the number of policyholders in the portfolio, can be efficiently handled. Finally, we discuss the sensitivity of the price to model parameters and study the accuracy of our approach to standard pricing methods. We emphasize that the resolution of the system of non-linear PDEs by neural networks presented in this paper is general and is not limited to life insurance.

We would also like to point out that BSDEs with jumps, which we apply in this paper to represent the solution to our system of non-linear PDEs, are usually used to represent solutions to non-linear partial integro-differential equations with Lévy jump measures. Recently, \cite{castro2021deep} considered a neural network to solve a PIDE in a similar fashion to ours, in the sense that an additional network is trained for the jump component. However, our system of non-linear PDEs and our jump components are different from the single PIDE in \cite{castro2021deep}.

The paper is structured as follows. In Section \ref{fininsmodel}, we present in detail the financial and insurance model. Section \ref{pricingsection} derives the pricing PDEs, discusses their representation as a BSDE with jumps and the numerical implementation by neural networks. In Section \ref{numresults}, we present numerical results and Section \ref{conclusion} provides concluding remarks.

\section{Financial and Insurance Model}\label{fininsmodel}

Throughout the paper, we work on a probability space $(\Omega ,{\mathbb{F}},\mathbb{P})$ equipped
with a filtration ${\mathbb{F}}=(\mathcal{F}_{t})_{0\leq t\leq T}$ and a
finite time horizon $T<\infty $. We consider a general financial market consisting of a bank account, a bond and an equity. We use a two-factor interest rate model to describe the short-rate process and Heston model to describe the dynamics of the equity price and its volatility. The insurance risk is modeled by a jump process with stochastic force of mortality.

\subsection{The financial market}
We assume that the dynamics of the risk-free rate under an equivalent martingale measure $\mathbb{Q}$ is given by the G2++ two-additive-factor gaussian model:
\beqo
r(t)&=&\psi(t)+x(t)+y(t),\quad r(0)=r_0,\\
dx(t)&=&-ax(t)dt+\sigma_xdW_1^{\mathbb{Q}}(t),\quad x(0)=0,\\
dy(t)&=&-by(t)dt+\sigma_ydW_2^{\mathbb{Q}}(t),\quad y(0)=0,
\eeqo
where $\psi$ is a deterministic function such that the initial term structure predicted by the model is fitted to the market term structure. The G2++ model is equivalent to two-factor Hull-White model under appropriate re-parametrization (see e.g. Section 4.2.5 in \cite{brigo2007interest}). For hedging in incomplete markets, models need to be defined under the real-world measure $\mathbb{P}$. We follow the approach from \cite{ait2010estimating} and introduce the dynamics
\beqo
r(t)&=&\psi(t)+x(t)+y(t),\quad r(0)=r_0,\\
dx(t)&=&\big(\delta_x \sigma_x-ax(t)\big)dt+\sigma_xdW_1(t),\quad x(0)=0,\\
dy(t)&=&\big(\delta_y \sigma_y-by(t)\big)dt+\sigma_ydW_2(t),\quad y(0)=0,
\eeqo
where we use constant risk premiums $\delta_x, \delta_y$ to switch from the martingale measure to the real-world measure.

The dynamics of the risk-free bank account is given by
\beqo
dR(t)=R(t)r(t,x(t),y(t))dt,\quad R(0)=1,
\eeqo
where we emphasize that the risk-free rate $r$ depends on $(t,x(t),y(t))$. In our G2++ model, the dynamics of the bond price $P(t)=\be^\bq[e^{-\int_t^{T^*}r(s)ds}|\mathcal{F}_t]$, which matures at time $T^\ast\geq T$, is given by
\beqo
dP(t)=P(t)\Big(\big(r(t,x(t),y(t))+\zeta(t)\big)dt+A(t)dW_1(t)+B(t)dW_2(t)\Big),
\eeqo
for some given deterministic functions $A, B$ (see e.g. Section 4.2.3 in \cite{brigo2007interest}):
\begin{align*}
A(t)=-\sigma_x \frac{1-e^{-a(T^\ast-t)}}{a},\quad B(t)=-\sigma_y \frac{1-e^{-b(T^\ast-t)}}{b},
\end{align*}
and the risk premium $\zeta$:
\beqo
\zeta(t)=A(t)\delta_x+B(t)\delta_y.
\eeqo
The risk premium $\zeta$ should be positive, hence the risk premiums $\delta_x, \delta_y$ should be negative (as always, we assume that $\sigma_x, \sigma_y$ are positive).

We assume that the equity follows the Heston stochastic volatility model (\cite{heston1993closed}):
\beqo
dS(t)&=&S(t)\Big(\big(r(t,x(t),y(t))+\gamma\sqrt{v(t)}\big)dt+\sqrt{v(t)}dW_3(t)\Big),\quad S(0)=1,\\
dv(t)&=&k(\eta-v(t))dt+\sigma_v\sqrt{v(t)}dW_4(t),\quad v(0)=v_0,
\eeqo
where we also assume a constant risk premium $\gamma$ for the equity. We assume that $v_0>0$, $k>0$ and $2k\eta\geq \sigma_v^2$ so that $v$ is strictly positive.

The policyholder pays a single premium for an equity-linked life insurance policy. The initial premium is invested in the policyholder's account value and the policyholder's funds are invested in the bond and the equity. The policyholder's account dynamics is then given by
\beqo
dF(t)&=&uF(t)\Big(\big(r(t,x(t),y(t))+\zeta(t)\big)dt+A(t)dW_1(t)+B(t)dW_2(t)\Big)\\
&&+(1-u)F(t)\Big(\big(r(t,x(t),y(t))+\gamma\sqrt{v(t)}\big)dt+\sqrt{v(t)}dW_3(t)\Big)-cF(t)dt,\\
F(0)&=&F_0,
\eeqo
where $F_0$ is the initial premium paid by the policyholder, $u$ is a fixed percentage of the account invested in the bond, and $c$ is a constant fee deducted by the insurer to cover the financial guarantees and insurance protection embedded in the contract. The portfolio of equity-linked contracts, which we introduce in the next section, is homogeneous in the sense that all policyholders invest the same single premium, follow the same investment strategy and have the same guarantees and insurance protection, i.e. $F_0$, $u$ and $c$ are equal and fixed for all policyholders. We point out that the value of the fee $c$ results from pricing the contract. In this paper we assume that the fee $c$ has already been set by a pricing actuary and we calculate the price of the contracts given $c$ . However, we can also find $c$ such that the price of the portfolio of equity-linked contracts is zero and we can call such a fee - a fair fee.

Finally, all Brownian motions involved in the interest rate and the equity  models are correlated with $dW_{i}(t)dW_{j}(t)=\rho_{ij} dt$ for $i,j=1,\ldots,4$.

\subsection{The insurance portfolio}

The portfolio consists of $n$ policies. All policyholders have the same age and are entitled to two types of benefits: a death benefit $D$ paid at the moment that the insured dies (provided he dies in $[0,T]$) and a survival benefit $S$ paid at terminal time $T$ if the insured survives that time. The benefits are contingent on the values of the policyholder's account $F$, hence we will use the notations $D(t,F(t))$ and $S(F(T))$ for death and survival benefits, respectively.

We assume that the lifetimes of the policyholders $(\tau_k)_{k=1,...,n}$ at policy inception are, conditional on the mortality intensity, independent and
identically distributed with mortality intensity $\lambda$, i.e. we assume that
    \beqo\label{exponentiallaw}
    {\mathbb{P}}\big(\tau_k>t|(\lambda(s),0\leq s \leq t)\big)=e^{-\int_0^t\lambda(s)ds}.
    \eeqo
The spot mortality intensity at calendar time $t$ of a head aged $x$ at time 0 is denoted by $\lambda (t)$. Following \cite{luciano2008modelling} and \cite{luciano2012delta}, we assume that $\lambda (t)$ follows the Feller process without mean reversion:
\begin{equation}
d\lambda(t)=q\lambda(t)dt+\sigma_\lambda \sqrt{\lambda(t)}dW_5(t),\quad \lambda(0)=\lambda_0,\label{lambdadyn}
\end{equation}
where $\lambda_0>0$ and $q>0$. The mortality intensity can attain zero value but \cite{luciano2008bis} showed that the probability of such an event is negligible for calibrated mortality data. The choice of this model is motivated by its parsimony-few parameters to be calibrated and its ability to fit cohort life tables due to the lack of mean reversion. Moreover, under the dynamics \eqref{lambdadyn}, the survival probability can be obtained in the closed-form:
\beq
{\mathbb{P}}\big(\tau_k>t\big)=\mathbb{E}\left[e^{-\int_0^t\lambda(s)ds}\right]=e^{\beta(t)\lambda(0)},\label{survivalfunction}
\eeq
with $\beta(t)=\frac{1-e^{bt}}{b_1+b_2e^{bt}},b=-\sqrt{q^2+2\sigma^2_\lambda},b_1=\frac{b+q}{2},b_2=\frac{b-q}{2}$.

The lives of the policyholders are dependent since they are all affected by the path of $\lambda (t)$ (systematic mortality risk). However, conditionally to the sample path followed by $\lambda (t)$, the lives are independent (unsystematic mortality risk). Similar as in credit risk, we can compute the probability that two policyholders survive (see e.g. Section 9.5 in \cite{mcneil2005quantitative}). Using the conditional independence, we obtain:
\begin{align*}
 {\mathbb{P}}\left(\tau_i>T,\tau_j>T\right)&=\mathbb{E}\left[ {\mathbb{P}}\big(\tau_i>T,\tau_j>T|(\lambda(s),0\leq s \leq T)\big)  \right]\\
 &= \mathbb{E}\left[ {\mathbb{P}}\big(\tau_i>T|(\lambda(s),0\leq s \leq T)\big) \times {\mathbb{P}}\big(\tau_j>T|(\lambda(s),0\leq s \leq T)\big) \right]\\
 &=\mathbb{E}\left[e^{-2 \int_0^T\lambda(s)ds}\right]=e^{\tilde{\beta}(T)\lambda(0)},\quad i\neq j,
\end{align*}
where $\tilde{\beta}(T)=\frac{2\left(1-e^{\tilde{b}T}\right)}{\tilde{b}_1+\tilde{b}_2e^{\tilde{b}T}},\tilde{b}=-\sqrt{q^2+4\sigma^2_\lambda},\tilde{b}_1=\frac{\tilde{b}+q}{2},\tilde{b}_2=\frac{\tilde{b}-q}{2}$, which is an increasing function of $\sigma_\lambda$. Therefore, the higher the volatility of the mortality intensity, the higher the systematic risk that both policyholders survive.

We introduce the processes $N$ and $J$ which count the number of deaths and in-force policies in the insurance portfolio, respectively:
\begin{align*}
N(t)&=\sum_{k=1}^n\mathbf{1}\{\tau_k\leq t\},\quad N(0)=0,\\
J(t)&=n-N(t),\quad J(0)=n.
\end{align*}
In addition, the compensated counting process:
\beq
\tilde{N}(t)=N(t)-\int_0^t(n-N(s-))\lambda(s)ds,\label{Ntildedynamics}
\eeq
is a martingale, and it will be used in further derivations. Finally, in the remainder of the paper, we assume that the processes which drive the insurance risk $(N,\lambda)$ are independent of the financial market.

The financial and insurance models considered in this paper are very popular in actuarial applications. Let us remark that different models can also be used and the methods presented below still hold.

\section{Pricing equity-linked life insurance}\label{pricingsection}

Even though the insurer can invest in a bank account, a bond or an equity, the insurance liabilities are not perfectly hedgeable due to the presence of the stochastic volatility risk and the independent actuarial risks. Therefore, there is no unique pricing technique for insurance contracts and one needs to disentangle the hedgeable and non-hedgeable part of the liability. For instance, \cite{bayraktar2009valuation} used the instantaneous Sharpe ratio and \cite{liang2017indifference} considered the principle of equivalent utility. In this paper, we follow the approach of \cite{delongfvtheory} and use the fair valuation operator with the instantaneous standard deviation risk margin. Our method for pricing can be described in the following steps:
\begin{enumerate}
	\item[\textbf{1.}] We set up a hedging portfolio composed of a bond, an equity and a bank account. To find the optimal investments, we minimize the local variance (quadratic variation) of the net asset value process, which is defined as the excess value of the hedging portfolio over the fair price of the liability. In case of a complete market, the minimized local variance is zero. However, the incompleteness of the market leads to a residual risk, as measured by the remaining local variance.
\item [\textbf{2.}] The insurer asks a compensation for this residual non-hedgeable risk and we assume that it is priced via an instanteneous standard deviation risk margin. \item [\textbf{3.}] The price can then be expressed as the solution to a system of non-linear PDEs. The system of PDEs describes the prices given the number of in-force policies and the number of equations is equal to the number of policies in the portfolio,
	\item[\textbf{4.}] In order to solve these PDEs, we reformulate the PDEs as a forward-backward stochastic differential equation with jumps and approximate the price using deep neural networks.
\end{enumerate}

\subsection{Fair pricing with the instantaneous standard deviation risk margin}

We assume that the insurer can invest in the bank account, the bond and the equity in order to hedge the financial guarantees and the benefits embedded in the contracts. Let $\theta_1$ denote the amount of money invested in the bond, and $\theta_2$ denote the amount of money invested in the equity. The remaining amount is invested in the bank account. We assume that $\theta_1$ and $\theta_2$ are Markov control strategies and are functions of $(t,x(t),y(t),F(t),v(t),\lambda(t),J(t))$. Let $V^\theta:=(V^\theta(t),0\leq t\leq T)$ denote the self-financing hedging portfolio under a strategy $\theta=(\theta_1,\theta_2)$. The dynamics of the insurer's hedging portfolio is given by
\beqo
dV^\theta(t)&=&\theta_1(t)\Big(\big(r(t,x(t),y(t))+\zeta(t)\big)dt+A(t)dW_1(t)+B(t)dW_2(t)\big)\\
&&+\theta_2(t)\Big(\big(r(t,x(t),y(t))+\gamma\sqrt{v(t)}\big)dt+\sqrt{v(t)}dW_3(t)\big)\\
&&+(V^\theta(t)-\theta_1(t)-\theta_2(t))r(t,x(t),y(t))dt+J(t-)cF(t)dt-D(t,F(t))dN(t),\\
V^\theta(0)&=&nF(0).
\eeqo
Moreover, let us remark that the survival benefits $J(T)S(F(T))$ are paid at the terminal time $T$ and therefore, do not affect the insurer's hedging portfolio.

In order to simplify the derivations, we introduce the 5-dimensional It\^{o} process $Z(t)=(x(t),y(t),F(t),v(t),\lambda(t))$ which models the five risk factors:
\beq
dZ(t)=\mu(t,Z(t))dt+\sigma(t,Z(t))dW(t),\label{dynamicsZ}
\eeq
with \begin{equation}
\mu(t,Z(t))=\begin{pmatrix}
\delta_x\sigma_x-ax(t)\\
\delta_y\sigma_y-by(t)\\
\left(r(t,x(t),y(t))-c+u\zeta(t)+(1-u)\gamma\sqrt{v(t)}\right)F(t)\\
k(\eta-v(t))\\
q\lambda(t)
\end{pmatrix},
\end{equation}
and \beq
\sigma(t,Z(t))=\begin{pmatrix}
	\sigma_x & 0 & \ldots & \ldots & 0 \\
	0 & \sigma_y & 0 & \ldots & 0 \\
	uF(t)A(t) & u F(t) B(t) & (1-u) F(t) \sqrt{v(t)} & 0 & 0 \\
	0 & \ldots & 0 & \sigma_v\sqrt{v(t)} & 0 \\
	0 & \ldots & \ldots & 0 & \sigma_\lambda \sqrt{\lambda(t)}
\end{pmatrix}.
\eeq
In the expression \eqref{dynamicsZ}, $W(t)$ is a 5-dimensional correlated Brownian motion with correlation matrix $\mathcal{Q}$, i.e. $dW_{i}(t)dW_{j}(t)=\rho_{ij} dt$ where $\rho_{ij} $ is the $(i,j)$-component of $\mathcal{Q}$ (for $i,j=1,\ldots,5$).

Let $\varphi:=(\varphi(t),0\leq t\leq T)$ denote the continuous-time fair valuation operator representing the fair price of the liabilities. We assume that $\varphi$ is a function of $(t, x(t),y(t),F(t),v(t),\lambda(t),J(t))$ and is denoted by $\varphi^{J(t)}(t,Z(t))$. We introduce the process $X(t)=\varphi^{J(t)}(t,Z(t))-V^\theta(t)$ representing the difference between the price of the liabilities and the value of the insurer's hedging portfolio. Consequently, the process $NAV(s) =-X(s)$ determines the excess of the assets over the technical provision, and is called the net asset value. Let $\Theta(s,Z(s))= (\theta_1(s)A(s),\theta_1(s)B(s),\theta_2(s)\sqrt{v(s)},0,0)$ denote a 5-dimensional vector related to the hedging strategy. By application of the multidimensional It\^{o}'s lemma, we find the dynamics:
\begin{eqnarray*}
&& dX(s)= \Big\{X(s)r(s,x(s),y(s))+\varphi_{t}^{J(s-)}(s,Z(s)) +\nabla \varphi^{J(s-)}(s,Z(s)) \cdot  \mu(s,Z(s)) -cJ(s-)F(s)\\
&& -\varphi^{J(s-)}(s,Z(s))r(s,x(s),y(s))+ \Big( \varphi^{J(s-)-1}(s,Z(s))+D(s,F(s))- \varphi^{J(s-)}(s,Z(s))  \Big) J(s-)\lambda(s)   \\
&& - \theta_1(s) \zeta(s) - \theta_2(s)\gamma\sqrt{v(s)}\Big\}ds\\
&& +\frac{1}{2}\Tr\Big( \sigma(s,Z(s)) \mathcal{Q} \sigma^\intercal(s,Z(s)) \Hessian_z \varphi^{J(s-)}(s,Z(s)) \Big) ds \\
&&+\Big( \varphi^{J(s-)-1}(s,Z(s))+D(s,F(s))- \varphi^{J(s-)}(s,Z(s))  \Big)d\tilde{N}(s)\\
&&+\Big( \nabla \varphi^{J(s-)}(s,Z(s)) \sigma(s,Z(s)) -\Theta(s,Z(s)) \Big) dW(s).
\end{eqnarray*}
Here, $\nabla  \varphi $ and $\Hessian_z \varphi $ denote the gradient and the Hessian of the function $ \varphi$ with respect to $Z$, $\Tr$ denotes the trace of a matrix.

We are interested in finding the valuation operator $\varphi$ which satisfies the property:
\beq\label{pricing_limit}
\lefteqn{\lim_{h\rightarrow 0}\left\{\frac{\be\Big[X(t+h)e^{-\int_t^{t+h}r(s)ds}-X(t)\Big|Z(t)=z,J(t)=k\Big]}{h}\right.}\nonumber\\
&&\left.+\frac{RM\Big[X(t+h)e^{-\int_t^{t+h}r(s)ds}-X(t)\Big|Z(t)=z,J(t)=k\Big]}{h}\right\}=0.
\eeq
As the one-period risk margin, we consider the standard deviation:
\beqo
\lefteqn{RM\Big[X(t+h)e^{-\int_t^{t+h}r(s)ds}-X(t)\Big|Z(t)=z,J(t)=k\Big]}\\
&=&\frac{1}{2}\alpha \sqrt{Var\Big[X(t+h)e^{-\int_t^{t+h}r(s)ds}-X(t)\Big|Z(t)=z,J(t)=k\Big]}\sqrt{h}.
\eeqo
The parameter $\alpha$ can be interpreted as the insurer's risk aversion coefficient towards the non-hedgeable risks.

We can choose the hedging strategy $\theta$ so that the insurer does not charge a risk margin for the hedgeable financial risk. The optimal investments in the bond and the equity are then determined by minimizing the instantaneous (local) variance of the net asset value process. Therefore, we need to compute the quadratic variation $\left[X\right]_s$ and find the bivariate process $(\theta_1(s),\theta_2(s))  $ which minimizes this quantity. The quadratic variation is given by
\beq
&&\left[X\right]_T=\int_0^T\underbrace{\Big(\nabla \varphi(s,Z(s)) \sigma(s,Z(s)) -\Theta(s,Z(s)) \Big) \mathcal{Q} \Big(\nabla \varphi(s,Z(s)) \sigma(s,Z(s)) -\Theta(s,Z(s)) \Big) ^\intercal }_{g(s,\theta_1(s),\theta_2(s))}ds \nonumber\\
&&+\Big( \varphi^{J(s-)-1}(s,Z(s))+D(s,F(s))- \varphi^{J(s-)}(s,Z(s))  \Big)dN(s).\label{xprocess}
\eeq
The optimal hedging strategy is then obtained by solving the first-order conditions:
\begin{align*}
\frac{\partial g(s,\theta_1,\theta_2)}{\partial \theta_1}&=-2\sum_{j=1}^{4} \Big(\rho_{1j}A(s) +\rho_{2j}B(s) \Big) \Big(\nabla \varphi(s,Z(s)) \sigma(s,Z(s)) -\Theta(s,Z(s)) \Big)_{j}\\
&=0,\\
\frac{\partial g(s,\theta_1,\theta_2)}{\partial \theta_2}&=-2\sum_{j=1}^{4}\rho_{3j} \sqrt{v(s)}\Big(\nabla \varphi(s,Z(s)) \sigma(s,Z(s)) -\Theta(s,Z(s)) \Big)_{j}\\
&=0.
\end{align*}
We remark that the index $j=5$ does not appear in the derivatives since the mortality intensity is not correlated with the financial risk factors. The first-order conditions yield a system of two equations with two unknowns for which the solutions are given by
\beqo
\theta^\ast_1(t,Z(t))&=& u \varphi_f(t,Z(t)) F(t)\\
&&+\frac{\varphi_x(t,Z(t)) \sigma_x \rho^x(t)+ \varphi_y(t,Z(t)) \sigma_y \rho^y(t)+\varphi_v (t,Z(t))\sigma_v \sqrt{v(t)} \rho^v(t)}{A(t)\rho^x(t)+B(t)\rho^y(t)},\\
\theta^\ast_2(t,Z(t))&=& (1-u) \varphi_f(t,Z(t)) F(t)\\
&& + \frac{\varphi_x(t,Z(t)) \sigma_x}{\sqrt{v(t)}} \left(\rho_{31}-\left(\rho_{31}A(t)+\rho_{32}B(t)\right)\frac{\rho^x(t)}{A(t)\rho^x(t)+B(t)\rho^y(t)} \right)\\
&&+\frac{\varphi_y(t,Z(t)) \sigma_y}{\sqrt{v(t)}} \left(\rho_{32}-\left(\rho_{31}A(t)+\rho_{32}B(t)\right)\frac{\rho^y(t)}{A(t)\rho^x(t)+B(t)\rho^y(t)} \right)\\
&&+\varphi_v(t,Z(t)) \sigma_v \left(\rho_{34}-\left(\rho_{31}A(t)+\rho_{32}B(t)\right)\frac{\rho^v(t)}{A(t)\rho^x(t)+B(t)\rho^y(t)} \right),
\eeqo
with some deterministic functions, which take into account the correlation between the different risk factors:
\begin{align*}
\rho^x(t)&=A(t)+\rho_{12}B(t)-\rho_{31}\rho_{13}A(t)-\rho_{31}\rho_{23}B(t),\\
\rho^y(t)&=B(t)+\rho_{12}A(t)-\rho_{32}\rho_{13}A(t)-\rho_{32}\rho_{23}B(t),\\
\rho^v(t)&=\rho_{14}A(t)+\rho_{24}B(t)-\rho_{34}\rho_{13}A(t)-\rho_{34}\rho_{23}B(t).
\end{align*}
 As it is usually the case, it turns out that the optimal hedging strategy appears as a delta-hedging strategy involving the sensitivities of the liabilities to the underlying financial risk factors and adjusted for the dependence structure.

Plugging the optimal parameters into the quadratic variation \eqref{xprocess} leads to
\beqo
&&\left[X\right]^\ast_T=\int_0^T\Big(\nabla \varphi(s,Z(s)) \sigma(s,Z(s)) -\Theta^\ast(s,Z(s)) \Big) \mathcal{Q} \Big(\nabla \varphi(s,Z(s)) \sigma(s,Z(s)) -\Theta^\ast(s,Z(s)) \Big) ^\intercal ds \nonumber\\
&&+\Big( \varphi^{J(s-)-1}(s,Z(s))+D(s,F(s))- \varphi^{J(s-)}(s,Z(s))  \Big)dN(s),
\eeqo
which represents the remaining quadratic variation. We remark that the vector $\Big(\nabla \varphi \sigma -\Theta^\ast \Big)$ can be written as $\Big(\nabla \varphi \sigma^\ast \Big)$ for an approriate matrix $\sigma^\ast$.\footnote{Given the optimal parameters, it is easy to decompose $\Theta^\ast$ as $\Theta^\ast=\nabla \varphi\tilde{\sigma}$. Then the appropriate matrix is $\sigma^\ast=\sigma-\tilde{\sigma}$ since $\nabla \varphi(\sigma-\tilde{\sigma})=\nabla \varphi \sigma -\Theta^\ast$.} Similarly, the risk-adjusted drift $\mu^\ast$ can be obtained from
\begin{equation*}
	\nabla \varphi^{J(s-)}(s,Z(s)) \cdot  \mu^\ast(s,Z(s)) = \nabla \varphi^{J(s-)}(s,Z(s)) \cdot  \mu (s,Z(s)) - \theta_1^\ast (s) \zeta(s,x(s),y(s)) - \theta_2^\ast(s)\gamma\sqrt{v(s)}.
\end{equation*}
If we consider the gradient component related to $f$, we find that
\begin{align*}
\varphi_f^{J(s-)}(s,Z(s))  \mu^{\ast,f}(s,Z(s))&=\varphi_f^{J(s-)}(s,Z(s))\left[\left(r(s,x(s),y(s))-c+u\zeta(s)+(1-u)\gamma\sqrt{v(s)}\right)F(s)\right]\\
&-\varphi_f^{J(s-)}(s,Z(s))uF(s)\zeta(s) \\
&-\varphi_f^{J(s-)}(s,Z(s))(1-u)F(s)\gamma\sqrt{v(s)}\\
&=\varphi_f^{J(s-)}(s,Z(s))\left(r(s,x(s),y(s))-c\right).
\end{align*}
Hence, the risk-adjusted drift of the fund $F(t)$ is the risk-free rate $r(t,x(t),y(t))$ minus the fee rate $c$, and the equity risk-premium is cancelled. 

Under the optimal hedging strategy, we find that
\begin{align}
\lefteqn{\lim_{h\rightarrow 0} \frac{\mathbb{E} \big[X(t+h)e^{-r(t)h}-X(t)|Z(t)=z,J(t)=k\big]}{h}}\nonumber\\
&=&\varphi _{t}^{k}(t,z)+\nabla \varphi^{k}(t,z) \cdot \mu^\ast(t,z)+\frac{1}{2}\Tr\Big( \sigma(t,z) \mathcal{Q} \sigma^\intercal(t,z) \Hessian_z \varphi^{k}(t,z) \Big) \notag\\
&&-ckf+\Big(\varphi^{k-1}(t,z)+D(t,f)-\varphi^{k}(t,z) \Big)k\lambda-\varphi^{k}(t,z)r(t,x,y),\label{exp}
\end{align}%
and
\begin{align}
&\lim_{h\rightarrow 0} \frac{\text{Var}\big[X(t+h)e^{-r(t)h}-X(t)|Z(t)=z,J(t)=k\big]}{h}\notag\\
&=\Big( \nabla \varphi^{k}(t,z) \sigma^\ast(t,z) \Big)\mathcal{Q} \Big( \nabla \varphi^{k}(t,z) \sigma^\ast(t,z) \Big)^\intercal +\Big(\varphi^{k-1}(t,z)+D(t,f)-\varphi^{k}(t,z) \Big)^2 k\lambda. \label{var}
\end{align}%
The last quantity represents the remaining quadratic variation and quantifies the non-hedgeable risks of the net asset value process $X$. We can conclude that the continuous-time fair valuation operator $\varphi$ solves the following system of non-linear PDEs:
\begin{equation}\label{PDEsolution}
\left\{
\begin{aligned}
&\varphi _{t}^{k}(t,z)+\nabla \varphi^{k}(t,z) \cdot \mu^\ast(t,z)+\frac{1}{2}\Tr\Big( \sigma(t,z) \mathcal{Q} \sigma(t,z)^\intercal \Hessian_z \varphi^{k}(t,z) \Big) \\
&-ckf+\Big(\varphi^{k-1}(t,z)+D(t,f)-\varphi^{k}(t,z) \Big)k\lambda(t)-\varphi^{k}(t,z)r(t,x,y)\\
&+\alpha\sqrt{\Big( \nabla \varphi^{k}(t,z) \sigma^\ast(t,z) \Big)\mathcal{Q} \Big( \nabla \varphi^{k}(t,z) \sigma^\ast(t,z) \Big)^\intercal +\Big(\varphi^{k-1}(t,z)+D(t,f)-\varphi^{k}(t,z) \Big)^2 k\lambda}=0,\\
&\varphi^{k}(T,z)=kS(f),	
\end{aligned}
\right.
\end{equation}
for $k\in\{0,\ldots,n\} $. For each $k$, we have to deal with a 5-dimensional PDE. The number of PDEs in the system \eqref{PDEsolution} is equal to $n+1$.

\subsection{The PDE, FBSDE and FSDE representations}

It is well known that there is a close relation between partial differential equations (PDEs) and backward stochastic differential equations (BSDEs). The solutions $(\varphi^k(t,z))_{k=0,...,n}$ to the PDEs \eqref{PDEsolution} satisfy the following forward-backward stochastic differential equation (FBSDE) with jumps (see e.g. \cite{barles1997backward} or Chapter 4 in \cite{delong2013backward}):
\begin{multline}\label{BSDErepresentation_zero}
\varphi^{J(t)}(t,Z(t))=J(T)S(F(T))\\
+\int_{t}^{T}\Upsilon^{J(s-)}\Big(s,Z(s),\varphi^{J(s-)}(s,Z(s)),\varphi^{J(s-)-1}(s,Z(s))\Big)ds+\int_t^TD(s,F(s))J(s-)\lambda(s)ds\\
-\int_{t}^{T} \Big( cJ(s-)F(s)+\varphi^{J(s-)}(s,Z(s)) r(s,x(s),y(s))\Big) ds\\
-\int_{t}^{T} \nabla \varphi^{J(s-)}(s,Z(s)) \sigma (s,Z(s)) dW(s)\\
-\int_{t}^{T} \Big( \varphi^{J(s-)-1}(s,Z(s))- \varphi^{J(s-)}(s,Z(s))  \Big)d\tilde{N}(s),
\end{multline}
\begin{eqnarray}\label{Forward_BSDErepresentation_zero}
Z(t)&=&Z(0)+\int_0^t\mu^*(s,Z(s))ds+\int_0^t\sigma(s,Z(s))dW(s),\nonumber\\
dJ(t)&=&-dN(t),
\end{eqnarray}
where
\begin{multline}\label{functionf}
\Upsilon^{J(s-)}\Big(s,Z(s),\varphi^{J(s-)}(s,Z(s)),\varphi^{J(s-)-1}(s,Z(s))\Big)\\
=\alpha\Big\{ \Big( \nabla \varphi^{J(s-)}(s,Z(s)) \sigma^\ast(s,Z(s)) \Big)\mathcal{Q} \Big( \nabla \varphi^{J(s-)}(s,Z(s)) \sigma^\ast(s,Z(s)) \Big)^\intercal \\ +\Big(\varphi^{J(s-)-1}(s,Z(s))+D(s,F(s))-\varphi^{J(s-)}(s,Z(s)) \Big)^2 J(s-)\lambda(s) \Big\}^{1/2}.
\end{multline}
The equation \eqref{BSDErepresentation_zero} is a one-dimensional BSDE with jumps where the generator and the terminal condition depend on a process satisfying a forward stochastic differential equation. In our case, the FSDE is a 6-dimensional equation given by \eqref{Forward_BSDErepresentation_zero}, where five components are diffusions and one component is driven by a step process. The BSDE \eqref{BSDErepresentation_zero} can also be written as a FSDE:
\begin{multline}\label{BSDErepresentation}
\varphi^{J(t)}(t,Z(t))=\varphi^{n}(0,Z(0))\\
-\int_{0}^{t}\Upsilon^{J(s-)}\Big(s,Z(s),\varphi^{J(s-)}(s,Z(s)),\varphi^{J(s-)-1}(s,Z(s))\Big) ds-\int_0^tD(s,F(s))J(s-)\lambda(s)ds\\
+\int_{0}^{t} \Big( cJ(s-)F(s)+\varphi^{J(s-)}(s,Z(s)) r(s,x(s),y(s))\Big) ds\\
+\int_{0}^{t} \nabla \varphi^{J(s-)}(s,Z(s)) \sigma (s,Z(s)) dW(s)\\
+\int_{0}^{t} \Big( \varphi^{J(s-)-1}(s,Z(s))- \varphi^{J(s-)}(s,Z(s))  \Big)d\tilde{N}(s).
\end{multline}
Starting from an initial price of the liabilities: $\varphi^{n}(0,Z(0))$ (i.e. for $n$ policyholders at time $t=0$ with initial values of the risk factors: $z=Z(0)$), the FSDE \eqref{BSDErepresentation} describes the price dynamics of the insurance liabilities through time.

The function $\Upsilon$ in \eqref{BSDErepresentation} represents the price for the non-hedgeable risks: the first term in \eqref{functionf} is the residual risk from the diffusion component (interest rate, volatility, equity and systematic mortality risks) and the second term in \eqref{functionf} is the residual risk from the jump component (unsystematic mortality risk). We can notice that the risk margin for the non-hedgeable financial risks takes into account the correlation matrix between these risks. The second term in \eqref{BSDErepresentation} represents the price for the expected death benefits. The first two terms in \eqref{BSDErepresentation} decrease the price as they are released over time. The next two terms increase the price. The third term stands for the continuous fees paid by the policyholders to cover the guarantees and benefits (since the fees to hedge the future liabilities are collected by the insurer, the price of the future liabilities must increase over time). The fourth term describes the effect of interest accrual and the time value of money on the price dynamics. The fifth term shows the impact of the diffusion component. Finally, the last term depicts the consequence of the death of a policyholder on the price (there is one policyholder less $J(s-) \mapsto J(s)= J(s-)-1 $ ).

In order to derive a numerical algorithm to determine the price at time 0, i.e $\varphi^{n}(0,Z(0))$, we treat $\varphi^{n}(0,Z(0))$ as parameters for the moment and view the FSDE \eqref{BSDErepresentation} as a way of computing the values of $\varphi^{J(t)}(t,Z(t))$ knowing the initial price $\varphi^{n}(0,Z(0))$ and the dynamics of $\varphi$. We consider a temporal discretization for the risk factors and the FSDE \eqref{BSDErepresentation}. For a set of time steps $0=t_0<t_1<\ldots<t_N=T$, we consider the simple Euler scheme for $n=1,\ldots,N-1$:
\beqo
Z(t_{n+1})-Z(t_{n})\approx \mu^\ast(t_n,Z(t_n))\Delta t_n+ \sigma(t_n,Z(t_n)) \Delta W(t_n),
\eeqo
for the diffusion process \eqref{dynamicsZ} and
\beqo
\tilde{N}(t_{n+1})-\tilde{N}(t_{n})=N(t_{n+1})-N(t_{n})-J(t_n)  \lambda(t_n) \Delta t_n,
\eeqo
for the compensated jump process \eqref{Ntildedynamics}. For the FSDE, the Euler scheme has the following form:
\begin{align}\label{BSDEdiscrete}
\begin{split}
&\varphi^{J(t_{n+1})}(t_{n+1},Z(t_{n+1}))=\varphi^{J(t_{n})}(t_{n},Z(t_{n}))\\
&-\alpha\Big\{ \Big( \nabla \varphi^{J(t_{n})}(t_{n},Z(t_{n})) \sigma^\ast(t_{n},Z(t_{n})) \Big)\mathcal{Q} \Big( \nabla \varphi^{J(t_{n})}(t_{n},Z(t_{n})) \sigma^\ast(t_{n},Z(t_{n})) \Big)^\intercal \\ &+\Big(\varphi^{J(t_{n})-1}(t_{n},Z(t_{n}))+D(t_{n},F(t_{n}))-\varphi^{J(t_{n})}(t_{n},Z(t_{n})) \Big)^2 J(t_{n})\lambda(t_{n}) \Big\}^{1/2} \Delta t_n\\
&-D(t_n,F(t_n))J(t_n)\lambda(t_n)\Delta t_n\\
&+ \Big( cJ(t_{n})F(t_{n})+\varphi^{J(t_{n})}(t_{n},Z(t_{n})) r(t_{n},x(t_{n}),y(t_{n}))\Big) \Delta t_n\\
&+\nabla\varphi^{J(t_{n})}(t_{n},Z(t_{n})) \sigma (t_{n},Z(t_{n}))\Delta W(t_n)\\
&+\Big( \varphi^{J(t_{n})-1}(t_{n},Z(t_{n}))- \varphi^{J(t_{n})}(t_{n},Z(t_{n}))\Big)  \Delta \tilde{N}(t_n)
\end{split}
\end{align}
where
\beqo
\Delta t_n=t_{n+1}-t_{n},\quad \Delta W(t_n)= W(t_{n+1})-W(t_{n}), \quad \Delta \tilde{N}(t_n)= \tilde{N}(t_{n+1})-\tilde{N}(t_{n}).
\eeqo
To move from time $t_n$ to time $t_{n+1}$, there are two unknown functions: the gradient of the price, $\nabla \varphi^{J(t_{n})}(t_{n},Z(t_{n}))$, and the price with one policyholder less, $\varphi^{J(t_{n})-1}(t_{n},Z(t_{n}))$. Following the work of \cite{weinan2017deep} and \cite{chan2018machine} (we remark that in both cases, the authors only consider BSDEs without jumps), we approximate both functions by the use of neural networks.

\subsection{Feed-forward neural networks as function approximators}\label{sectionNN}

Hereafter, we briefly introduce feed-forward neural networks (short \textit{neural networks}). In regression theory, networks can be seen as a broad generalization of generalized linear models (GLMs) where the linear predictor is replaced by a non-linear one.

Let us assume that the input is in dimension $d_{0}$ (the state variable $x$) and the output is in dimension $d_{1}$ (the number of value functions to estimate). The network is characterized by a number of layers $L+1 \in \mathbb{N} \backslash\{1,2\}$ with $m_{\ell}, \ell=0, \ldots, L,$ the number of neurons (units or nodes) on each layer: the first layer is the input layer with $m_{0}=d_{0}$, the last layer is the output layer with $m_{L}=d_{1},$ and the $L-1$ layers between are called hidden layers, where we choose for simplicity the same dimension $m_{\ell}=m, \ell=1, \ldots, L-1$.
A network is a function from $\mathbb{R}^{d_{0}}$ to $\mathbb{R}^{d_{1}}$ defined as the composition
$$
x \in \mathbb{R}^{d_0} \mapsto \mathcal{N}(x)=A_{L} \circ \varrho \circ A_{L-1} \circ \ldots \circ \varrho \circ A_{1}(x) \in \mathbb{R}^{d_1}
$$
Here, $A_{\ell}, \ell=1, \ldots, L$, are affine transformations represented by
$$
A_{\ell}(x)=\mathcal{W}_{\ell} x+\beta_{\ell},
$$
for a matrix of weights $\mathcal{W}_{\ell}$ and a bias term $\beta_{\ell}$. The non-linear function $\varrho: \mathbb{R} \rightarrow \mathbb{R}$ is called the activation function and is applied component-wise on the outputs of $A_{\ell},$ i.e., $\varrho\left(x_{1}, \ldots, x_{m}\right)=\left(\varrho\left(x_{1}\right), \ldots, \varrho\left(x_{m}\right)\right)$. Standard examples of activation functions are the sigmoid, the ReLU, the elu and tanh. For a general introduction on neural networks, we refer to \cite{goodfellow2016deep}.

The universal approximation theorem of \cite{hornik1989multilayer} states that networks can approximate any continuous function on a compact support arbitrarily well
if we allow for arbitrarily many neurons $q_{1} \in \mathbb{N}$ in the hidden layer.

\begin{figure}[tbp]
	\includegraphics[]{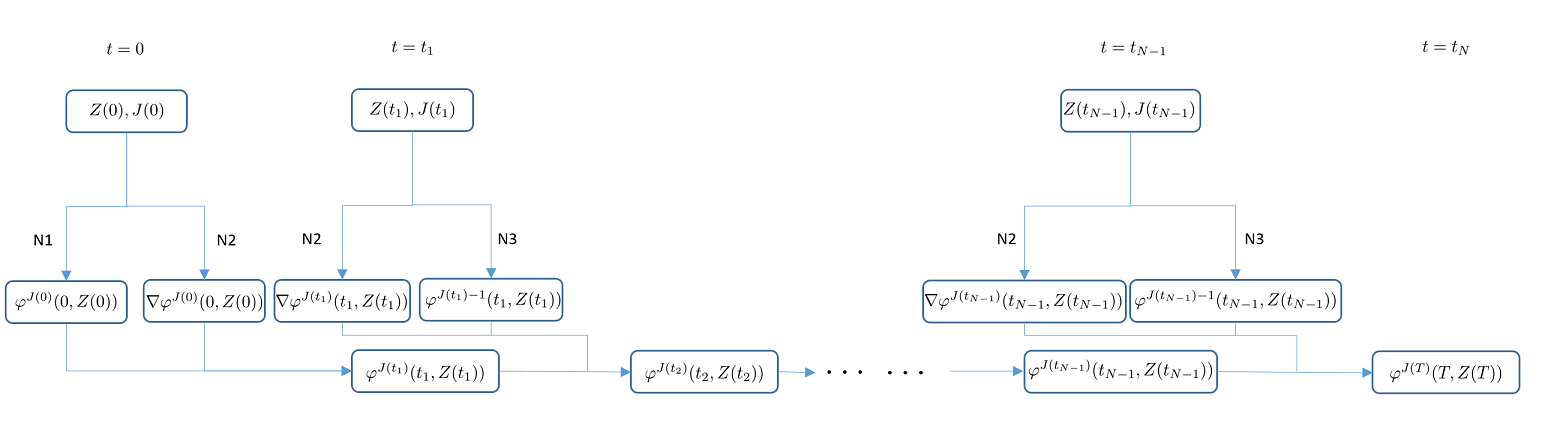}
	\caption{Illustration of the network architecture for solving the system of non-linear PDEs with $N$ time steps. The network N1 stands for the price at time 0, the network N2 approximates the gradient and the network N3 is related to the jump component and approximates the price with one policyholder less.}
	\label{graphnn}
\end{figure}

In the Euler scheme \eqref{BSDEdiscrete} of the FBSDE, we approximate the gradient and the price with one less policyholder by two respective neural networks. Moreover, in practice, prices need to be computed several times for different input parameters. Therefore, as suggested in \cite{weinan2017deep}, we determine the price at time 0 in a region of input parameters instead of a single space-point by including a third neural network. Overall, we have the three following networks as depicted in Figure \ref{graphnn}:
\begin{enumerate}
\item[\textbf{N1.}] The price at time 0, $\varphi^{J(0)}(0,Z(0))$, is approximated by a neural network $\mathcal{N}_1\ :(z,k)\in \mathbb{R}^6 \mapsto \varphi^{k}(0,z) \in \mathbb{R}$ with parameters $\phi$:
\begin{equation*}
\varphi^{J(0)}(0,Z(0))=\mathcal{N}^{\phi}_1(Z(0),J(0)).
\end{equation*}

\item[\textbf{N2.}] The gradient, $\nabla \varphi^{J(t_{n})}(t_{n},Z(t_{n}))$, is approximated by a second neural network $\mathcal{N}_2\ :(t,z,k)\in \mathbb{R}^7 \mapsto \nabla \varphi^{k}(t,z) \in \mathbb{R}^5$ with parameters $\chi$:
\begin{equation*}
\nabla \varphi^{J(t_{n})}(t_{n},Z(t_{n}))=\mathcal{N}^{\chi}_2(t_{n},Z(t_{n}),J(t_{n})).
\end{equation*}

\item[\textbf{N3.}] The price with one less policyholder, $\varphi^{J(t_{n})-1}(t_{n},Z(t_{n}))$, is approximated by a third neural network $\mathcal{N}_3\ :(t,z,k-1)\in \mathbb{R}^7 \mapsto \varphi^{k-1}(t,z) \in \mathbb{R}$ with parameters $\psi$:
\begin{equation*}
\varphi^{J(t_{n})-1}(t_{n},Z(t_{n}))=\mathcal{N}^{\psi}_3(t_{n},Z(t_{n}),J(t_{n})-1).
\end{equation*}
\end{enumerate}

By the use of the Euler scheme \eqref{BSDEdiscrete} and the neural networks, we can obtain a Monte-Carlo approximation $\widehat{\varphi}$ of $\varphi^{J(T)}(T,Z(T))$ which depends on the parameters of the neural networks $(\phi,\chi,\psi)$. These parameters are then estimated in order to minimize the quadratic loss function:
\begin{equation}\label{NN_objective}
	l(\phi,\chi,\psi)= \mathbb{E}\left[\Big|J(t_N)S(F(t_N))-\widehat{\varphi}^{J(t_{N})}(t_{N},Z(t_{N})) \Big|^2\right].
\end{equation}
The parameters can now be optimized by stochastic gradient descent-type (SGD) algorithm as it is usually done to train neural networks. In this paper, we use the Adaptive Moment Estimation (Adam) optimizer proposed by \cite{kingma2014adam} and the algorithm is implemented with the Keras R package (\cite{chollet2017kerasR}), which is a user-friendly API to TensorFlow.

\begin{rem}
\begin{itemize}
\item One can notice either from the PDEs \eqref{PDEsolution} or the FSDE \eqref{BSDErepresentation} that we face a recursive problem: in order to determine the price of the insurance liabilities for $n$ initial policyholders, we need to know the price for $k\in\{0,\ldots,n-1\} $. One possibility of implementation would have been to solve $n$ recursive optimizations with one neural network for the gradient in each optimization. Instead, we consider only one optimization with two neural networks (one for the gradient, one for the jump component) in order to reduce the computational cost of the problem.
\item Contrary to \cite{weinan2017deep}, we consider one single network for all time steps instead of one neural network per time step in order to reduce the numbers of parameters to be estimated. \cite{chan2018machine} showed that this neural network architecture significantly improves the precision of the results and the stability of the algorithms.

\end{itemize}
\end{rem}

\section{Numerical results}\label{numresults}

This section presents numerical results of our neural networks to price equity-linked contracts with death and survival benefits. In Section \ref{specification}, we briefly discuss our input parameters and the hyperparameters of the neural networks. In Section \ref{pricing1}, we solve the system of PDEs \eqref{PDEsolution} with our neural networks and analyse the impact of various model parameters on the price of equity-linked contracts. In Section \ref{pricing2}, we assess the accuracy of our approach for some specific contracts. Let us point out that we only investigate the initial price of the contracts, i.e. at time $t=0$.

\subsection{Specification of the insurance-financial model and neural networks}\label{specification}

For the financial and insurance model, we consider a realistic set of model parameters which are consistent with previous studies:
\begin{itemize}
	\item For the two-factor interest rate model, we consider
	\begin{equation*}
		a=0.2770,\ \sigma_x=0.0118,\ b=0.0551, \sigma_y=0.0136,
	\end{equation*}
	which are based on \cite{russo2019calibration}. Moreover, we chose: $\delta_x=\delta_y=-0.1,\psi(t)=0.02.$
	\item For the Heston stochastic volatility model, the parameters are
		\begin{equation*}
	\kappa=0.0231,\ \eta=0.9052, \sigma_v=0.1434,\gamma=0.0113,
	\end{equation*}
	which are consistent with \cite{li2016optimal}.
	\item For the force of mortality dynamics, we choose the parameters from \cite{luciano2008bis} which correspond to UK individuals aged 65 at time 0:
		\begin{equation*}
	q=0.11,\ \sigma_\lambda=0.007.
	\end{equation*}
	\item The correlation matrix for the interest rate factors, the equity, the volatility and the mortality intensity is given by
	\begin{equation}
	\mathcal{Q}=\begin{pmatrix}
	1 & -0.4 & 0.35 & 0 & 0 \\
	-0.4 & 1 & 0.08 & 0 & 0 \\
	0.35 & 0.08 & 1 & -0.3 & 0 \\
	0 & 0 & -0.3 & 1 & 0 \\
	0 & 0 & 0 & 0 & 1
	\end{pmatrix}
	\end{equation}
	which is consistent with \cite{grzelak2011affine}.	
\end{itemize}

Additional input parameters are presented in Table \ref{parameters}. We point out that the pricing neural network should not only deliver one price but a surface of prices which is a function of the interest rate factors, the premium, the volatility, the number of policyholders and the force of mortality at time $t=0$. Hence, we are interested in determining the price denoted by $\varphi^{J(0)}(0,Z(0))$. Here, $Z(0)=(x(0),y(0),F(0),v(0),\lambda(0))$ stands for the set of initial values for the interest rate factors, the premium, the volatility and the force of mortality respectively, and $J(0)$ stands for the number of policyholders at the inception. In our numerical example we investigate ranges of input parameters by changing the values for $v(0), F(0), \lambda(0), J(0)$ specified in Table \ref{parameters} by $+/-25\%$. For the interest rates, we consider the ranges $x(0)\in [0,0.01], y(0)\in [0,0.01]$.

\begin{table}[h]
	\centering
	\begin{tabular}{lrlr}
		\toprule
Parameter & Value & Parameter & Value \\
	\midrule
	$c$	        & 0.01  &$x(0)$ & $0$ \\
    $D^\star$	& 1.02    &$y(0)$ & $0$ \\
	$S^\star$	& 1.02   &$F(0)$ & $1$ \\
	$T$	& 1             &$v(0)$ & $0.1$ \\
	$\Delta t$	& 0.01   & $ \lambda(0)$ & $0.015$ \\
    $\alpha$ & 0.1        & $J(0)$ & $100$ \\
	\bottomrule
	\end{tabular}
	\caption{The base values of the initial parameters. \label{parameters}}
\end{table}

For the neural network architecture, different hyperparameters need to be chosen. Each of the neural networks \textbf{N1}, \textbf{N2}, \textbf{N3} defined in Section \ref{sectionNN} consists of two hidden layers of 20 neurons. For the activation function, we considered the ELU (exponential linear unit) which tends to accelerate the learning process (\cite{clevert2015fast}). Moreover, neural networks tend to have convergence issues when their inputs are not scaled and centered, and perform best when the inputs are in the range $\sim [-1,1]$. Therefore, we considered a min-max normalization of the inputs before passing them to the neural networks. Hence, we applied the function
\begin{equation*}
	g: \mathbb{R} \mapsto [-1,1]: g(x)=2 \times \left(\frac{x-\min(x,0)}{\max(x,0)-\min(x,0)} \right)-1.
\end{equation*}
Finally, we chose a batch size and a number of epochs equal to 200.\footnote{The computation time for the training of the price surface takes about 10 min on a computer with an Intel Core i7-4970MQ processor running at 2.90 GHz with 16 GB of RAM.} We considered 10000 sample paths for each risk factor and the default learning rate of the Adam optimizer. We tried different hyperparameters and we did not observed significant improvements in training of the neural networks.

\subsection{Pricing in the general case and sensitivity analysis}\label{pricing1}

Each equity-linked life insurance contract provides insurance death and survival benefits linked to the policyholder's account value. These financial guarantees ensure that the benefits are not lower than the minimal guarantee levels stipulated at the inception of the contract. We set the benefits:
\begin{align*}
D(t,F(t))&=(D^*-F(t))_+,\quad S(F(T))=(S^*-F(T))_+.
\end{align*}
The contract payoffs if the policyholder's account value goes below $D^*$ at the time of death or below $S^*$ in case of survival at the end of the contract. Hence, we are interested in the fair valuation of the financial guarantees embedded in the insurance contracts.

\subsubsection{Neural network's convergence diagnostics}

\begin{figure}[h]
	\centering
	\includegraphics[width=0.8\linewidth]{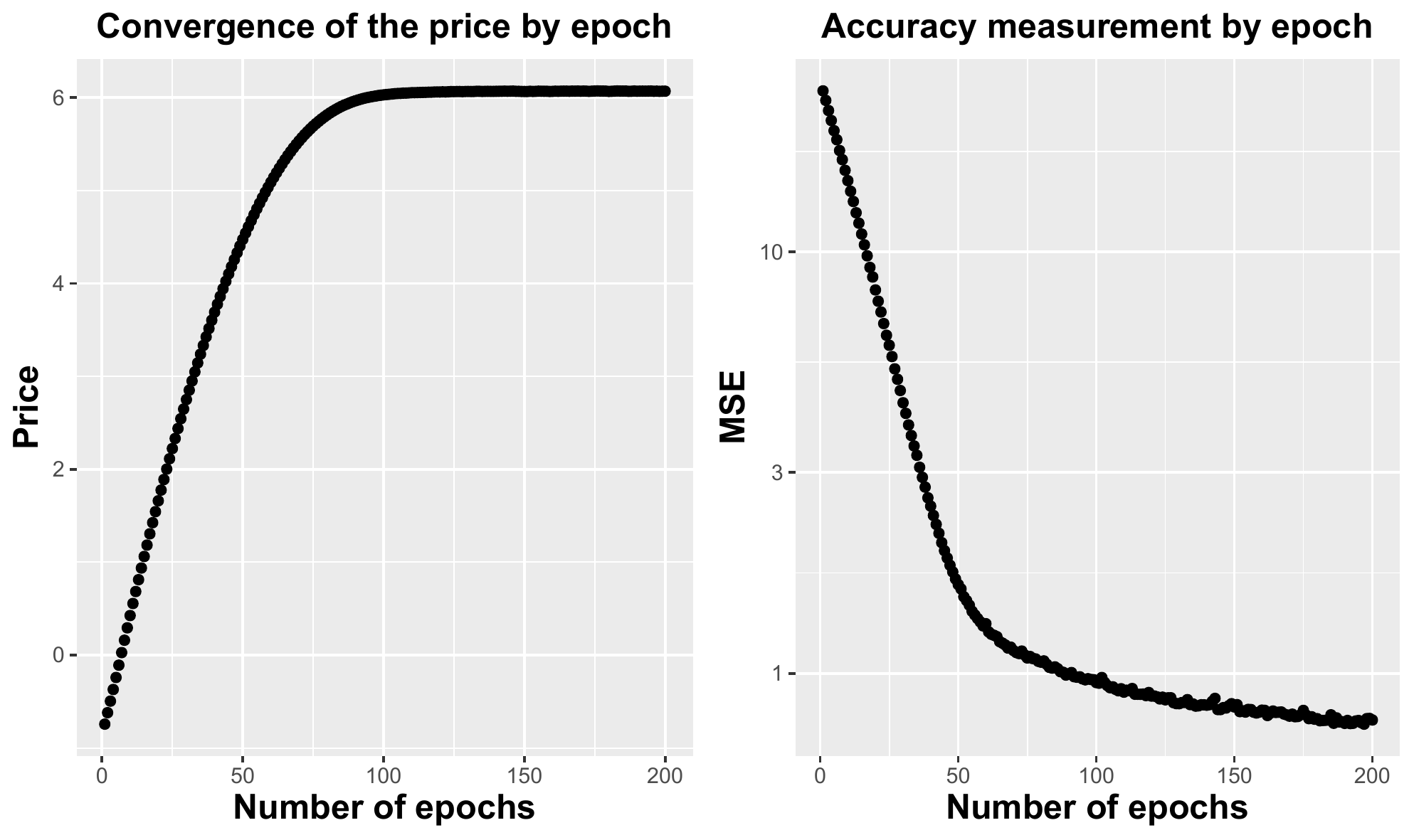}
	\caption{Diagnostics of the neural network training. On the left: Convergence of the price at time 0 per epoch. On the right: Mean Square Error per epoch.}
	\label{graph1}
\end{figure}

First, we consider the convergence of our algorithm for one price with the initial values $Z(0)=(0,0,1,0.1,0.015)$ and $J(0)=100$ policyholders from Table \ref{parameters}. Figure \ref{graph1} represents the convergence of the price and the mean square error against the number of epochs used for training the neural networks. Let us recall that our objective in the reinforcement learning is to minimize the mean square error \eqref{NN_objective}, hence we present the mean square error loss function to validate the convergence. We observe that, with the initialization of our algorithm by default, the price reaches a plateau after around 100 epochs. After 100 epochs, the MSE continues to slightly decreases but without significant impact on the price. We arrive at the price of the portfolio of the equity-linked contracts equal to $\varphi^{J(0)}(0,Z(0))=6.069$.

\begin{figure}[h]
	\centering
	\includegraphics[width=0.8\linewidth]{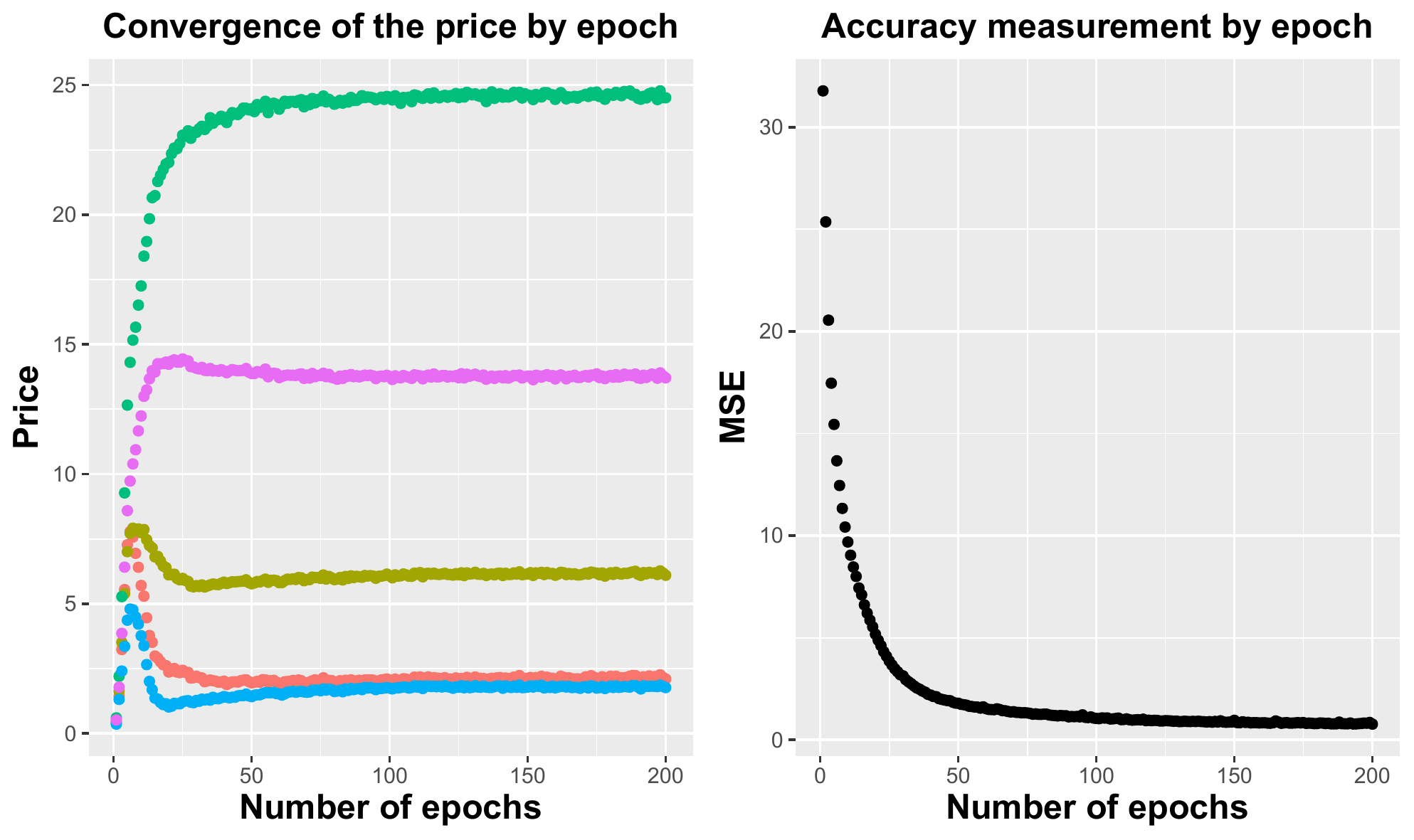}
	\caption{On the left: Convergence of the price at time 0 during the neural network training for five randomly chosen initial parameters. On the right: Mean Square Error per epoch during the neural network training of the surface of prices.}
	\label{graph2}
\end{figure}

Now, we study the convergence of the algorithm for a surface of prices for different sets of initial parameters $Z(0)$ and $J(0)$. The sets were constructed by taking the values given in Table \ref{parameters} and changing $v(0), F(0), \lambda(0), J(0)$ by $+/-25\%$ and considering $x(0)\in [0,0.01], y(0)\in [0,0.01]$. Figure \ref{graph2} shows the price convergence for 5 random choices of initial sets of parameters. We observe that the machine learns the price essentially in the first 100 epochs and then stabilizes. We remark that the price of the portfolio of the equity-linked contracts can possibly be negative as the contracts contain a fee rate. If the price is negative, it means that the fees deducted from the policyholder's account are too high in comparison with the guarantees embedded in the contract and the insurer expects to earn a profit.

From Figures \ref{graph1}-\ref{graph2} and similar analysis we performed, we can conclude that our algorithm is stable and the convergence is achieved.

\subsubsection{Price sensitivity analysis}

Hereafter, we illustrate the impact of various model parameters on the price of the portfolio of the equity-linked contracts. One of the purpose of this exercise is to validate that our PDEs provide a reasonable pricing operator.

\begin{figure}[h]
	\centering
	\includegraphics[width=0.8\linewidth]{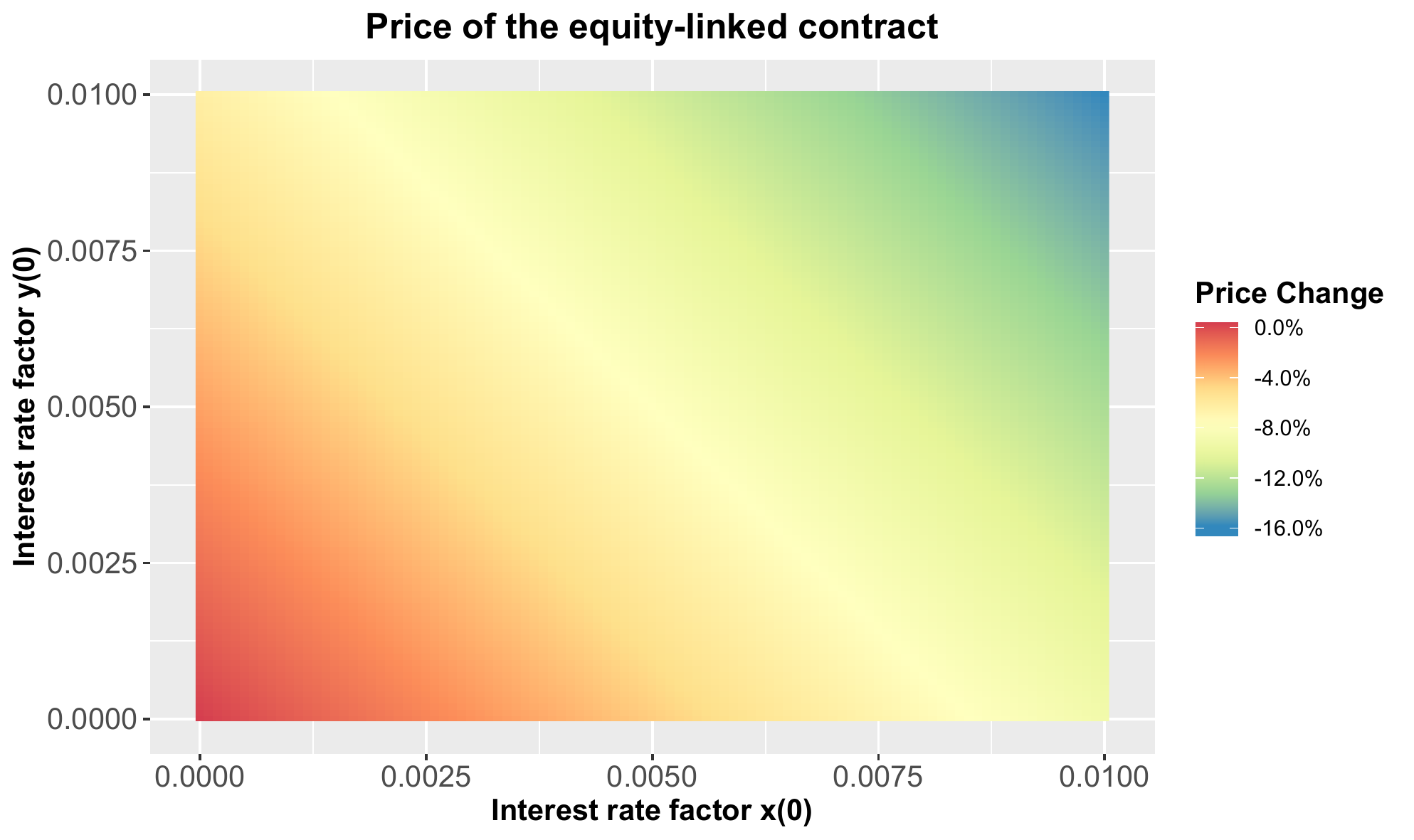}
	\caption{The price of the portfolio of equity-linked contracts for a range of the initial interest rate factors $x(0)\in \left[0;0.01\right]$ and $y(0)\in \left[0;0.01\right]$.}	\label{graphinterest}
\end{figure}

Figure \ref{graphinterest} represents the effect of the initial interest rate factors $x(0)$ and $y(0)$ on the price. As expected, the price decreases as the interest rate factors increase due to the time value of money. In particular, if both factors increase from $0$ to $0.01$, the price decreases by 16\%.

\begin{figure}[h]
	\centering
	\includegraphics[width=0.8\linewidth]{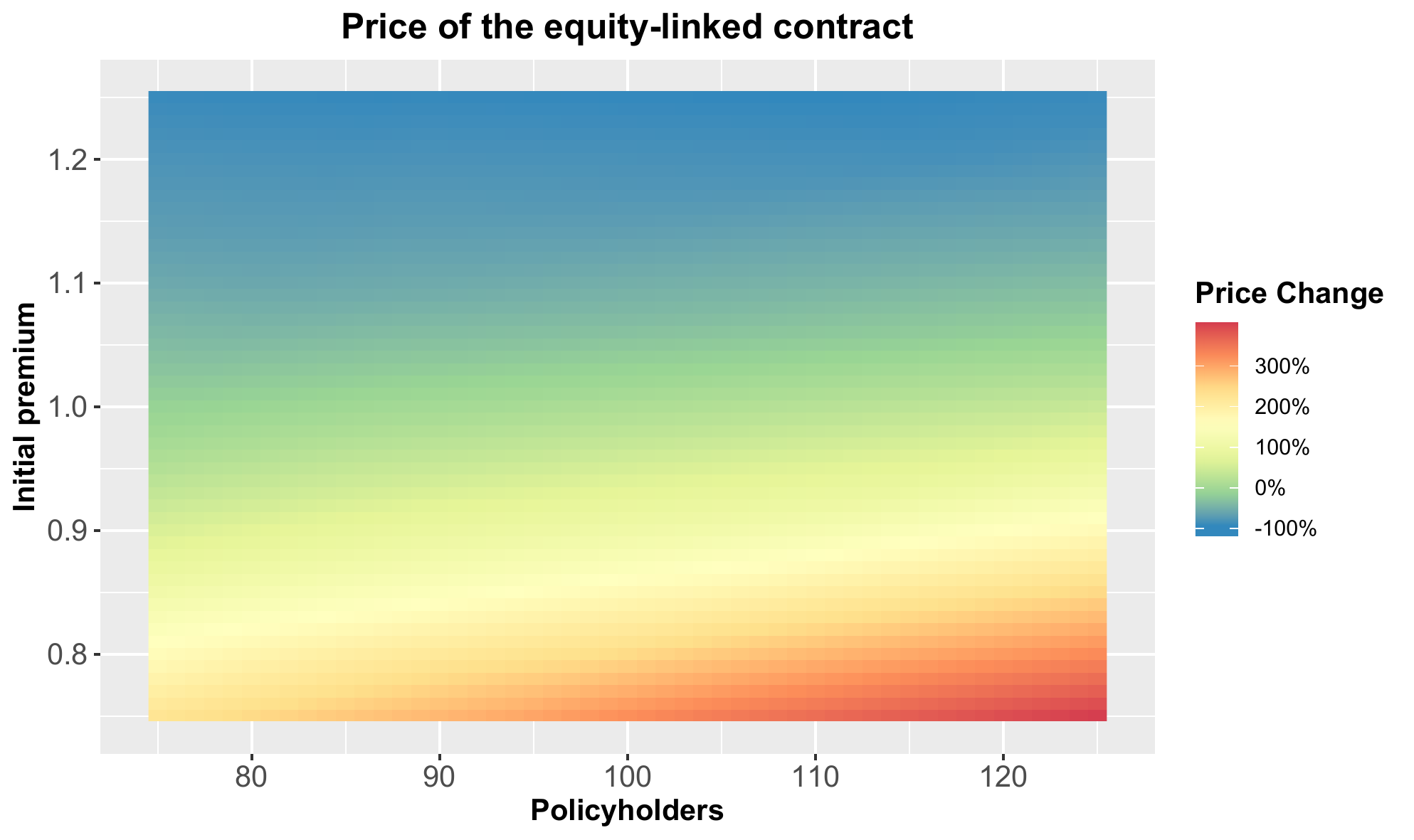}
	\caption{The price of the portfolio of equity-linked contracts for a range of the initial premiums $F(0)\in \left[0.75;1.25\right]$ and the numbers of policyholders $J(0)\in \left[75;125\right]$.}	\label{graph3}
\end{figure}

Figure \ref{graph3} shows the effect of varying the initial premium $F(0)$ and the number of policyholders $J(0)$ on the price. Obviously, the price of the portfolio of the equity-linked contracts increases with the number of policyholders since more policyholders mean more benefits to pay. Moreover, we note that the price of the portfolio of the equity-linked contracts decreases with the initial premium. Indeed, the increase of the initial fund value decreases the probability that the fund will go below the death guarantee level $D^*$ and the survival guarantee level $S^*$, and therefore decreases the price of the contract. We can observe that the price can double if the initial premium decreases by 10\% as depicted in Figure \ref{graph3}.

\begin{figure}[h]
	\centering
	\includegraphics[width=0.8\linewidth]{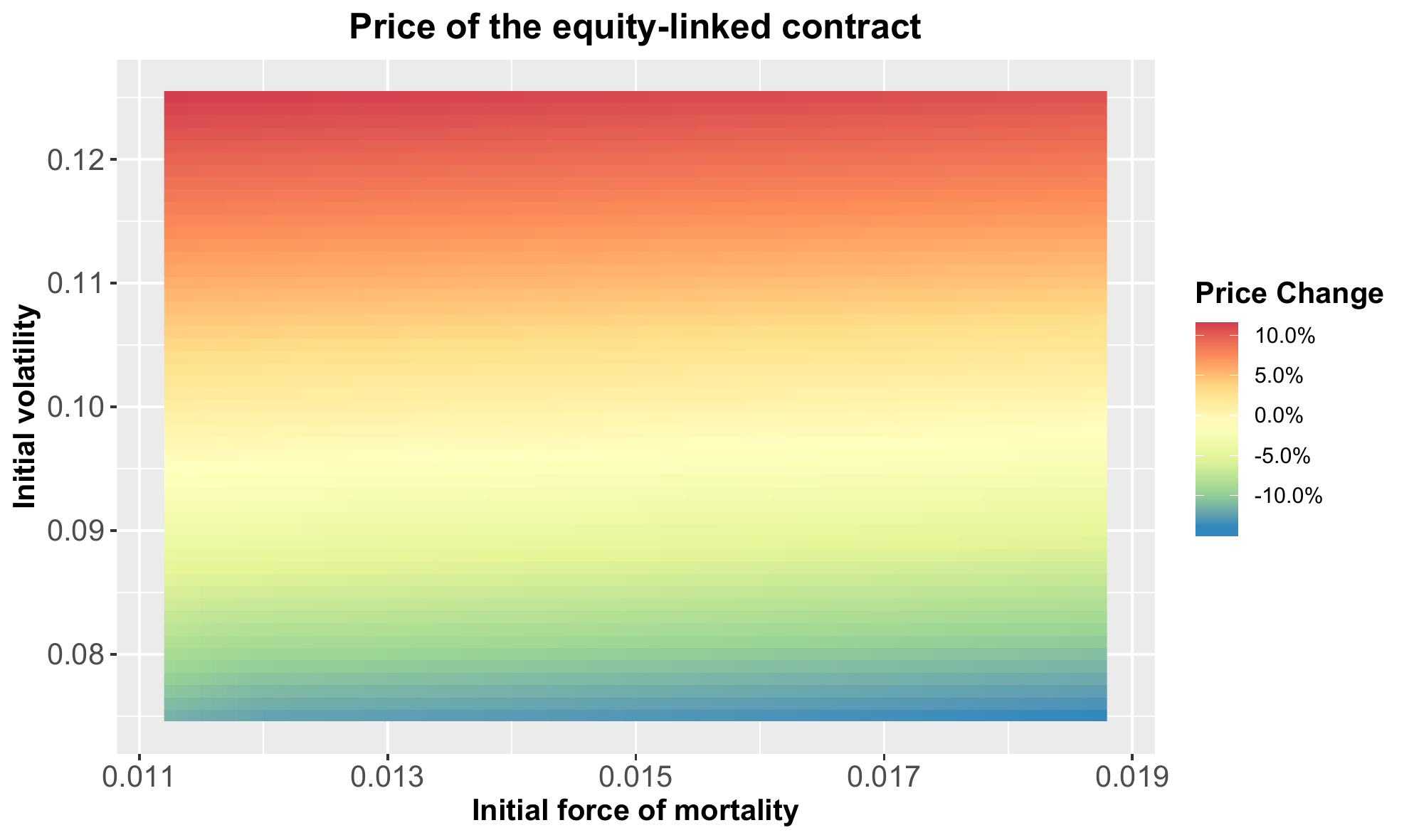}
	\caption{The price of the portfolio of equity-linked contracts for a range of the initial volatilities $v(0) \in \left[0.075;0.125\right]$ and the forces of mortality $ \lambda(0) \in \left[0.01125;0.01875\right]$.}
	\label{graph4}
\end{figure}

Figure \ref{graph4} depicts the impact of varying the initial volatility parameter $v(0)$ and the initial force of mortality $\lambda(0)$ on the price of the portfolio of the equity-linked contracts. As expected, we note that the price is an increasing function of the initial volatility. A highly volatile market environment implies that the stock and the fund values are less predictable, which will result in a higher risk charge required by the insurer in order to cover the guarantees. When the initial force of mortality increases, the value of the death benefit increases while the value of the survival benefit decreases as policyholders are more likely to die. On Figure \ref{graph4}, the price is a fairly constant function of the initial force of mortality, showing that the value reduction of the survival benefit compensates the value increase of the death benefit, at least for $D^*=S^*=1.02$ as in our numerical example.

We also investigate the impact of the correlation between the fund and the interest rate factor $x$ as well as the correlation between the fund and the volatility on the price of the liabilities. The results are presented in Table \ref{table:tabcorrelation}. As expected, the lower the correlation coefficient, the lower the price for the portfolio of the equity-linked contracts. This property reflects diversification effect between the financial risks and is consistent with the findings reported in \cite{gudkov2019pricing} and \cite{dai2015pricing}.

\begin{table}[h!]
	\begin{center}
		\begin{tabular}{lll}
			\toprule
			Change of the correlation &	\multicolumn{2}{c}{Change of the price}\\
			\cmidrule(ll){2-3}
			& $\rho_{xf} $ & $\rho_{fv} $\\
			\midrule
			$+0.1$ & +0.106\% & +0.131\% \\
			$-0.1$ & -0.248\% & -0.447\% \\
			\bottomrule
		\end{tabular}
		\caption{The percentage change of the price of the portfolio of equity-linked contracts for a correlation change by $\pm 0.1$.\label{table:tabcorrelation}}
	\end{center}
\end{table}

\begin{figure}[h]
	\centering
	\includegraphics[width=0.8\linewidth]{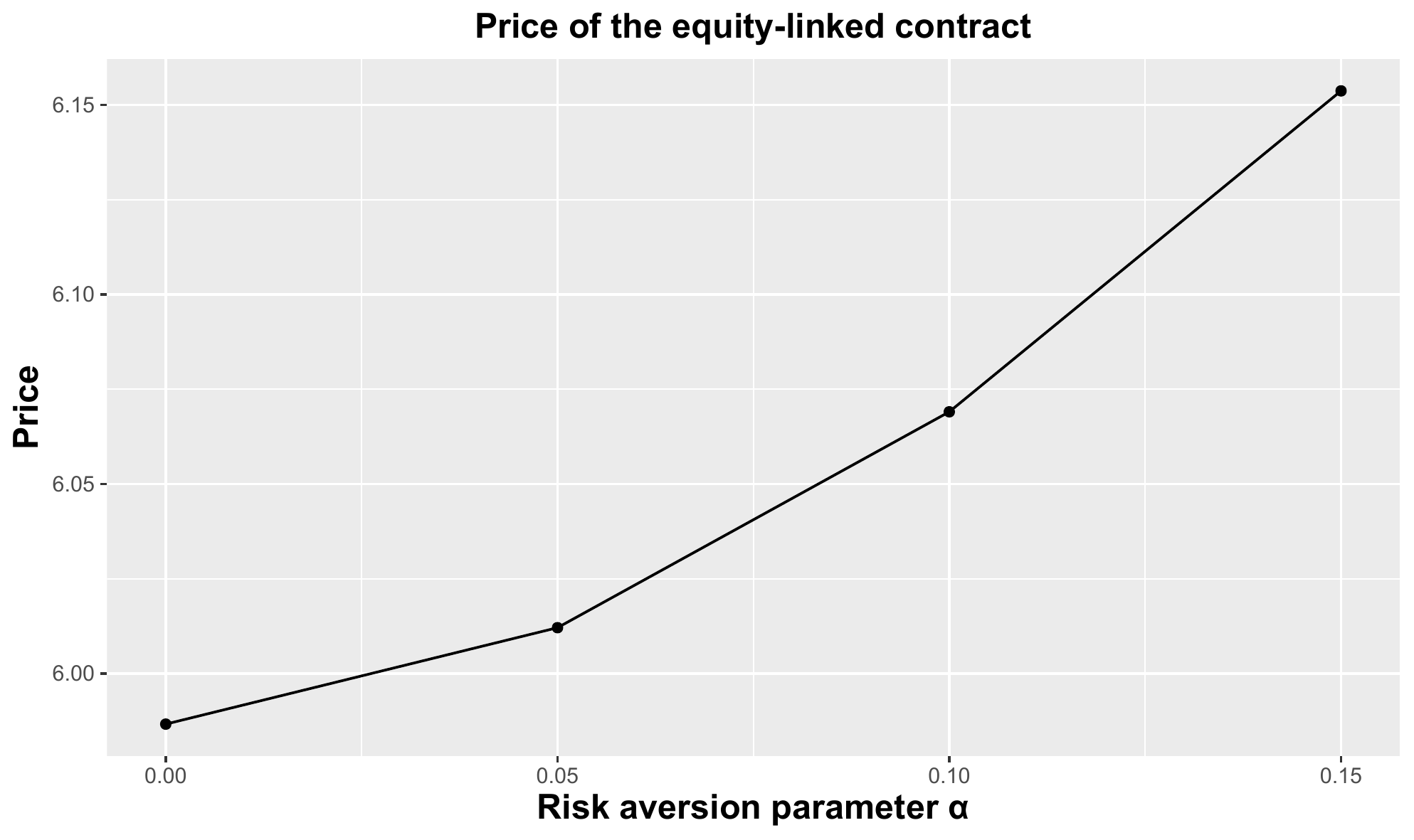}
	\caption{The price of the portfolio of equity-linked contracts in terms of the risk aversion parameter $\alpha$.}
	\label{graphriskaversion}
\end{figure}

Finally, we test the effect of the risk aversion coefficient $\alpha$. We expect that the more risk averse the insurer is, the larger the risk margin he will keep to cover the non-hedgeable risks and the larger the price of the portfolio of the equity-linked contracts. This is confirmed in Figure \ref{graphriskaversion}. We note that the relation between the price and the risk aversion $\alpha$ is not linear. The case $\alpha=0$ corresponds to the best estimate of the liability, which we discuss in the next subsection, when the insurer does not set any risk margin for the non-hedgeable risks.

\subsection{Numerical comparisons}\label{pricing2}

Hereafter, we assess the accuracy of our neural network approach by comparing it with Monte-Carlo pricing for two particular cases.

\subsubsection{Best Estimate Liability}

In insurance, the Best Estimate Liability (BEL) plays a central role in the valuation of insurance liabilities. It is defined as the expected present value of the future cashflows discounted by the risk-free rate. For our portfolio of the equity-linked life insurance contracts, the BEL is given by
\begin{eqnarray}\label{BE}
	BEL&=& \mathbb{E}\left[e^{-\int_{t}^{T}r(\tau)d\tau}J(T)S(F(T))
	+\int_{t}^{T}  e^{-\int_{t}^{s}r(\tau)d\tau} D(s,F(s))J(s)\lambda(s)ds\nonumber\right.\\
    &&\left.-\int_{t}^{T}e^{-\int_{t}^{s}r(\tau)d\tau} cJ(s)F(s)\Big|J(t),x(t),y(t),F(t),v(t),\lambda(t) \right]
\end{eqnarray}
where the first term is the expected survival benefits, the second term - the expected death benefits and the third term - the expected fees.
By the Feynman-Kac formula, the BEL is also solution to the following system of PDEs:
\begin{equation}\label{BEPDE}
\left\{
\begin{aligned}
&\varphi _{t}^{k}(t,z)+\nabla \varphi^{k}(t,z) \cdot \mu^*(t,z)+\frac{1}{2}\Tr\Big( \sigma(t,z) \mathcal{Q} \sigma(t,z)^\intercal \Hessian_z \varphi^{k}(t,z) \Big) \\
&-ckf+\Big(\varphi^{k-1}(t,z)+D(t,f)-\varphi^{k}(t,z) \Big)k\lambda(t)-\varphi^{k}(t,z)r(t,x,y)=0,\\
&\varphi^{k}(T,z)=kS(f),	
\end{aligned}
\right.
\end{equation}
for $k\in\{0,\ldots,n\} $.

\begin{figure}[h]
	\centering
	\includegraphics[width=0.9\linewidth]{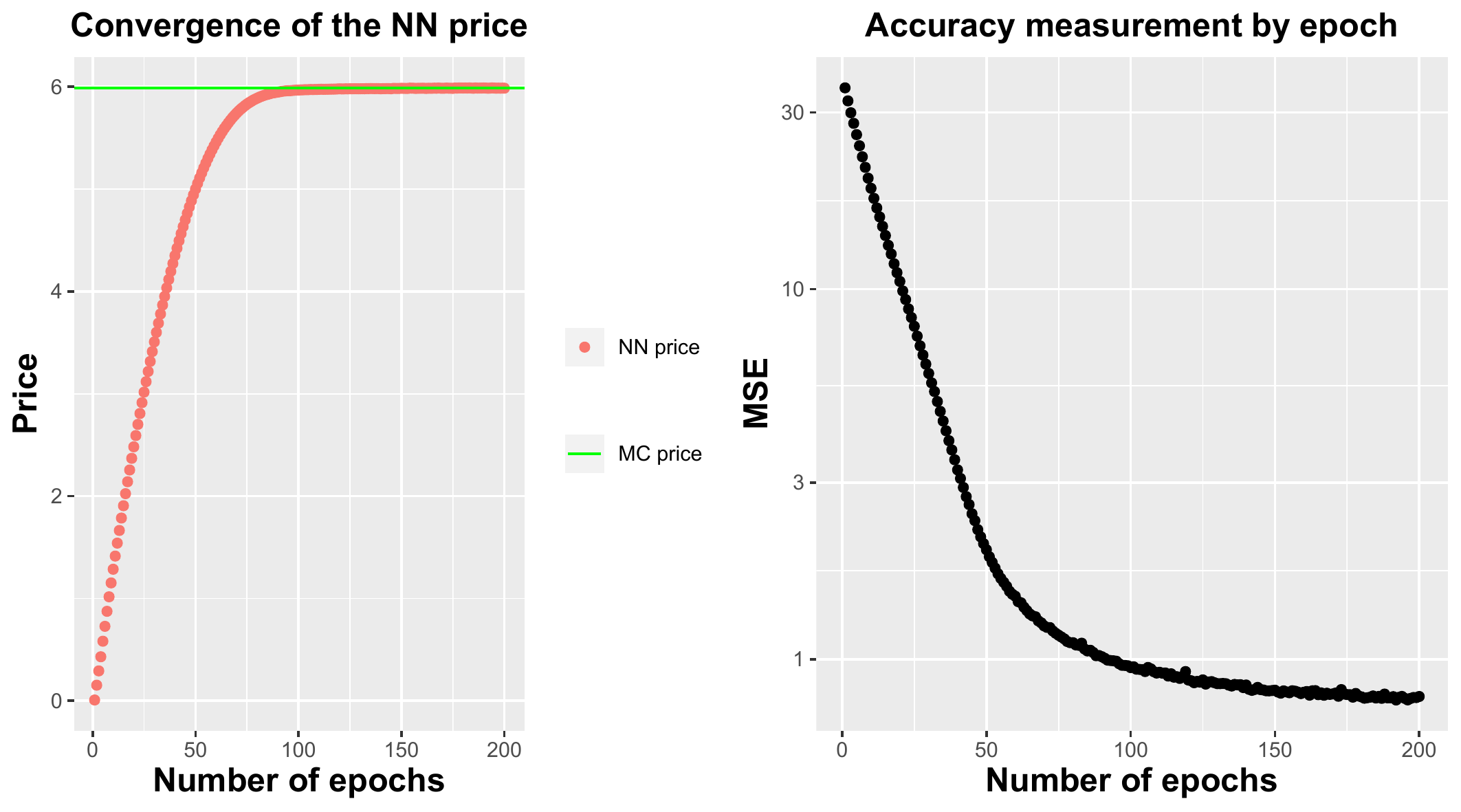}
	\caption{Neural Network price convergence to the Monte-Carlo price for the BEL and the Mean Square Error per epoch}
	\label{plotMC}
\end{figure}

The system \eqref{BEPDE} corresponds to our pricing equation \eqref{PDEsolution} without the risk margin (with $\alpha=0$). To assess the accuracy of our neural network, we compare the price obtained by an approximation of the BEL \eqref{BE} by 200.000 Monte-Carlo simulations and the price obtained by our neural network using the same parameters as in Table \ref{parameters}. Figure \ref{plotMC} shows that there is a good convergence of the neural network to the Monte-Carlo price. After 100 epochs, we obtain a relative error of c.a. $0.1\%$.

We could include the value of the BEL as an initial point from which we start to calibrate our neural networks for the price of the portfolio of unit-linked contracts since we expect that the true price with a risk margin for $\alpha>0$ should be close to the price with $\alpha=0$. Our tests showed that such an approach does not speed up the calibration of our neural networks mainly due to the fact that we still have to initiate the neural network for the gradient from a random starting point.

\subsubsection{Pricing a portfolio of GMMB contracts in the Black-Scholes framework}

In the following example, we consider a portfolio of equity-linked contracts in which only a survival benefit is guaranteed. Such contracts are often called Guaranteed Minimum Maturity Benefits (GMMB). We consider a simplified financial-actuarial market, namely the Black-Scholes financial market model and a jump process $J(t)$ with constant mortality $\lambda$ to model the deaths of policyholders. Under these assumptions, the pricing PDEs \eqref{PDEsolution} reduce to
\begin{equation}\label{PDEex2}
\left\{
\begin{aligned}
&\varphi _{t}^{k}(t,f)+\varphi_{f}^{k}(t,f)fr+\frac{1}{2} \varphi_{ff}^{k}(t,f)f^2\sigma_f^2-\varphi^{k}(t,f)r\\
&+\Big(\varphi^{k-1}(t,f)-\varphi^{k}(t,f) \Big) k \lambda \left(1-\frac{\alpha}{\sqrt{k\lambda}}
 \right)=0,\\
&\varphi^{k}(T,f)=kS(f),	
\end{aligned}
\right.
\end{equation}
for $k\in\{0,\ldots,n\} $.

We remark that this PDE was considered in (\cite{delongfvapplications}, Example 4.5). The first part of the PDE \eqref{PDEex2} exactly corresponds to the standard Black-Scholes PDE but the second part adds a risk margin for the uncertainty about the number of survivals which cannot be completely hedged. Applying the Feynman-Kac formula, the solution of the PDEs \eqref{PDEex2} is given by
\begin{equation}\label{MCprice}
\varphi^{n}(0,F(0))= \mathbb{E}^{\tilde{\mathbb{Q}}}\left[e^{-rT}J(T)S(F(T)) \right],
\end{equation}
where under $\tilde{\mathbb{Q}}$, the jump process has now a stochastic mortality intensity given by
\begin{equation*}
\tilde{\lambda}(t)=\lambda \left(1-\frac{\alpha}{\sqrt{J(t)\lambda}}\right).
\end{equation*}

We choose the parameters: $r=0.02,F(0)=1,\sigma_f=0.1,S^*=1.02,T=1,n=100, \alpha=0.1$. Under these parameters, $\tilde{\lambda}(t)$ is strictly positive which allows us to use a step process to simulate the number of deaths under the measure $\tilde{\mathbb{Q}}$.

\begin{figure}[h]
	\centering
	\includegraphics[width=0.8\linewidth]{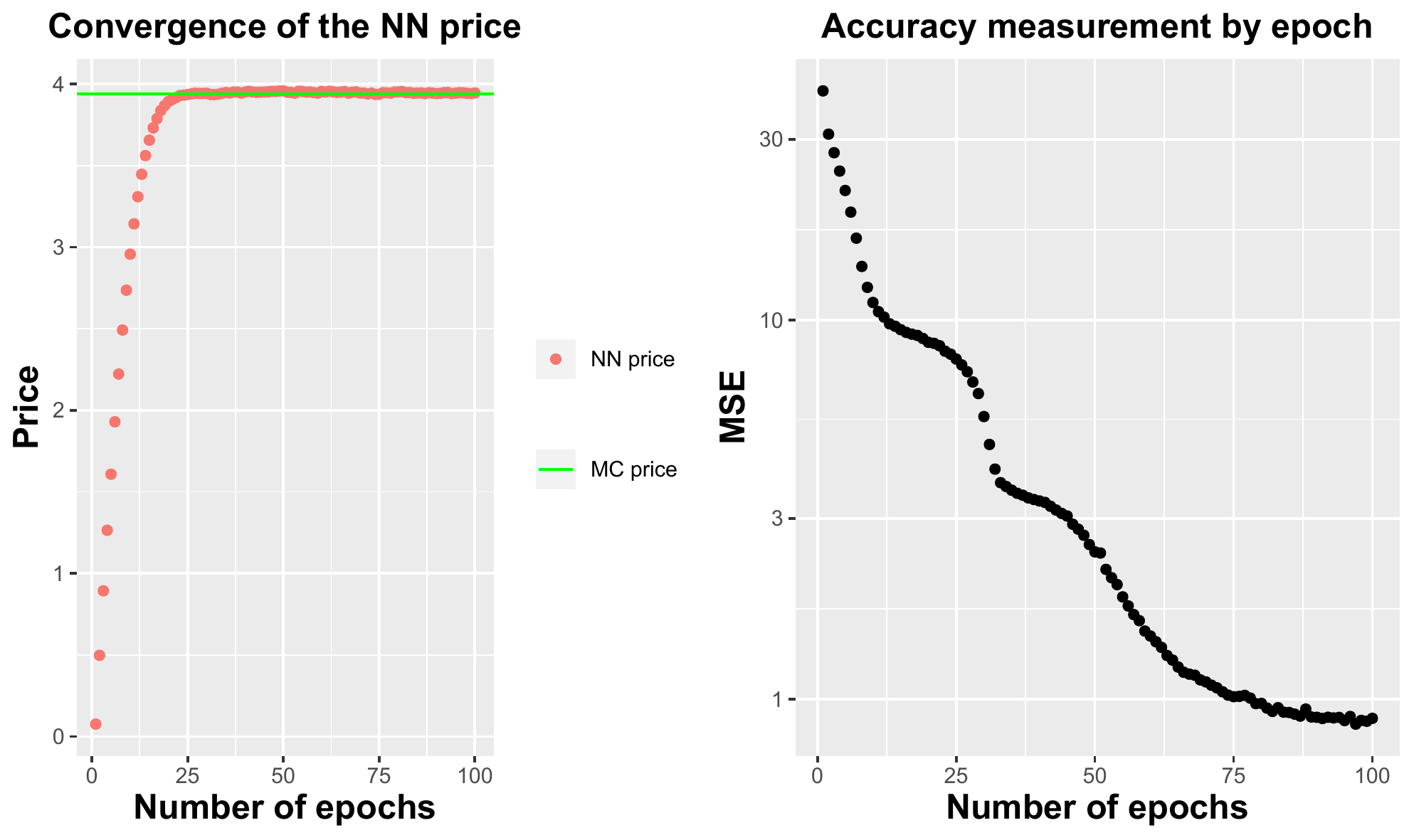}
	\caption{Neural Network price convergence to the Monte-Carlo price for the GMMB and Mean Square Error per epoch.}
	\label{plotMC2}
\end{figure}

Figure \ref{plotMC2} compares the Monte-Carlo price obtained by approximating \eqref{MCprice} with 200.000 simulations and the price obtained by neural networks by solving the PDEs \eqref{PDEex2}. We clearly observe a fast convergence after only 25 epochs and a relative error of c.a. $0.2\%$ after 100 epochs.

Summing up, both examples show that our algorithm is accurate. The remaining error is due to discretization steps applied in reinforcement learning, when we look for the optimal strategy in a discrete time, and Monte-Carlo simulations of discretized stochastic processes. Moreover, we note that our remaining MSE is of the order $[10^{-2};10^{-1}]$ which is consistent with orders obtained in \cite{chan2018machine}. The second example in this section also shows that our approach to handling the jump component performs very well.

\section{Conclusion}\label{conclusion}

In this paper, we have derived a partial differential equation for fair pricing of equity-linked life insurance contracts in a general financial-actuarial market with stochastic interest rate, equity, volatility and mortality. We assumed that the insurer can hedge its liabilities with a bank account, a bond and an equity and requires a compensation for non-hedgeable risks in the form of an instantaneous standard deviation risk margin. We derived a system of PDEs which described the fair valuation operator. We used the connection with BSDEs with jumps and proposed an efficient neural network architecture to solve the PDEs. The network consists essentially of two subnetworks: one to approximate the gradient component and one to approximate the jump component. We also included a third subnetwork for the price at time 0 which allows us to derive the price for a range of initial parameters rather than just one price.

Numerical results provided convergence diagnostics, price sensitivities to various model parameters and some comparisons with Monte-Carlo methods. In particular, we observed a fast convergence and stabilization of the price after a few dozens of epochs and a good approximation compared to prices obtained by Monte-Carlo simulations. In terms of sensitivity, we proved that the calibrated price agreed with intuition.

Our paper provides a stable computational algorithm for fair pricing of insurance liabilities in incomplete markets which can be used for a wide variety of financial and insurance products. Moreover, our neural network offers potential other applications in fields where a system of non-linear PDEs naturally arises, such as quantum mechanics, game theory, dynamic programming and many more.

\section{Acknowledgements}

The authors would like to thank the Editor and the anonymous referee who provided useful and detailed comments that substantially improved the current manuscript. Karim Barigou acknowledges the financial support of the Joint Research Initiative on ``Mortality Modeling and Surveillance" funded by AXA Research Fund.

\end{document}